\documentclass[twocolumn,showpacs,preprintnumbers,amsmath,amssymb,prl]{revtex4}%\documentclass[preprint,showpacs,preprintnumbers,amsmath,amssymb]{revtex4}
\usepackage{graphicx}% Include figure files
\usepackage{dcolumn}% Align table columns on decimal point
\usepackage{bm}% bold math

\def\cmo{CaMnO$\bf _3$}
\def\lmo{LaMnO$\bf _3$}
\def\lcmo{La$\bf _{1-x}$Ca$\bf _x$MnO$\bf _3$}
\def\range{$0 \le x \le 0.5$}

\begin{document}

%\preprint{Draft}
\title{Understanding the insulating phase in CMR manganites: Shortening of the Jahn-Teller long-bond across the phase diagram of \lcmo}

\author{E.~S.~Bo\v{z}in,$^1$ M.~Schmidt,$^{2}$ A.~J.~{DeConinck},$^1$  G.~Paglia,$^1$ J.~F.~Mitchell,$^3$
T.~Chatterji,$^4$ P.~G.~Radaelli,$^{2}$ {Th.}~Proffen,$^5$ and
S.~J.~L.~Billinge$^1$}

 \affiliation{$^1$Department of Physics and Astronomy, Michigan State
University, East Lansing, MI 48824-2320, USA}
 \affiliation{$^2$ISIS,
   CCLRC Rutherford Appleton Laboratory, Chilton-Didcot, OX11 0QX,
Oxfordshire, United Kingdom}
% \affiliation{$^3$Institute of Molecular
%Physics Polish Academy of Science, ul. Smoluchowskiego 17, 60-179
%Poznan, Poland}
 \affiliation{$^3$Materials Science Division, Argonne
National Laboratory, Argonne, Illinois 60439, USA}
 \affiliation{$^4$Institute Laue-Langevin, Bo\^{i}te Postale 156,
38042 Grenoble Cedex 9, France}
 %\affiliation{$^6$Department of
%Physics and Astronomy, University College London, London WC1E 6BT,
%United Kingdom}
 \affiliation{$^5$Lujan Neutron Scattering Center,
Los Alamos National Laboratory, Los Alamos, New Mexico 87545, USA}

\date{\today}

\begin{abstract}
The detailed evolution of the magnitude of the local Jahn-Teller
(JT) distortion in \lcmo\  is obtained across the phase diagram for
\range\  from high quality neutron diffraction data using the atomic
pair distribution function (PDF) method. A local JT distortion is
observed in the insulating phase for all Ca concentrations studied.
However, in contrast with earlier local structure studies its
magnitude is not constant, but {\it decreases} continuously  with
increasing Ca content. This observation is at odds with a simple
small-polaron picture for the insulating state.
\end{abstract}
\pacs{61.12-q,71.38.-k,75.47.Lx,75.47.Gk}
%}

\maketitle

Doped transition metal oxides are of fundamental interest for their
electronic properties that exhibit various types of colossal
responses such as high-temperature superconductivity, colossal
magnetoresistance (CMR) and ferroelectricity~\cite{milli;n98}, which
are still not fully understood.  A recent concept that may be
relevant for many of these systems is the observation of nanoscale
inhomogeneities that are thought to be intrinsic.  These can take
the form of nanoscale checkerboard-~\cite{komiy;prl05,hanag;n04} or
stripe-patterns~\cite{tranq;n95}, or less ordered
structures~\cite{fath;s99,uehar;n99,becke;prl02,loudo;prl06} that
are related to electronic phase separation.  Phase separation is
also found theoretically in computational models, which have been
used to explain the large changes in conductivity with temperature
and doping~\cite{dagot;s05}. However, the ubiquity and fundamental
importance of electronic inhomogeneities to the colossal effects is
not completely established~\cite{milwa;n05}. For example,
local-probe experiments where no electronic inhomogeneities are
observed have been made on samples that show a colossal
response~\cite{akiya;apl01,matsu;pc03} and some observed
inhomogeneities have been related to disorder in the chemical dopant
ions~\cite{mcelr;s05}.

One approach for establishing a correlation between inhomogeneities
and electronic properties is to study the complete phase diagram of
interesting systems.  We set out to examine the local structure of
the \lcmo\  (LCMO)  manganite family.  This is an archetypal system
for such studies because of the strong electron-phonon coupling,
through the JT effect, resulting in a large structural response to
electronic phase separation.  We probe this using the atomic pair
distribution function (PDF) measured from neutron powder diffraction
data. The signature of the metallic phase in local structural probes
is the absence of a JT long-bond~\cite{billi;prl96,booth;prl98}.
When the charges are localized in a polaronic insulating phase the
JT long-bond is
seen~\cite{billi;prl96,radae;prb96i,louca;prb97,booth;prl98,billi;prb00}.
A two phase fit of an undistorted and distorted phase to the low-r
region of the PDF can thus yield a quantitative measure of the phase
fraction of each phase as a function of $T$ and $x$ in the system
\lcmo~\cite{proff;apa01ii}. Unexpectedly, we found that the {\it
length} of the Jahn-Teller long-bond in the insulating phase
decreases continuously with increasing doping in this system,
contrary to the canonical
understanding~\cite{dagot;prpl01,louca;prb97,massa;jmmm01,billi;prb00}.
Here we report for the first time extensive, high real-space
resolution PDF data as a function of temperature and doping from $0
\le x \le 0.5$ on well characterized samples in the \lcmo\ system
using new high resolution, high-throughput, time-of-flight neutron
diffractometers. We rule out the existence of fully JT distorted
long bonds at 2.16~\AA\ in the CMR compositions that would be
expected in a simple single-site small-polaron scenario for the
phase above $T_c$.  The smooth continuous decrease in the length of
the JT long-bond is most easily explained in a delocalized or large
polaron picture.

\begin{figure}
\includegraphics[angle=270,width=0.45\textwidth]{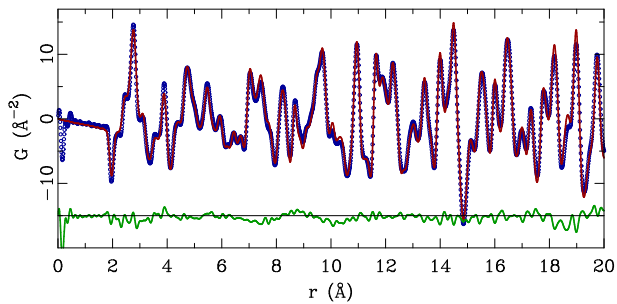}
\caption{\label{fig:grdata} Open circles are the $G(r)$ of $x=0.22$
sample at 10~K, solid red line is the calculated $G(r)$ from the
crystal structure model. The difference curve is shown offset below.}
\end{figure}

A series of 13 powdered LCMO samples with compositions spanning $0
\le x \le 0.5$ range were prepared using standard solid state
synthesis methods, and annealed to ensure oxygen stoichiometry~\cite{dabro;jssc89}
and characterized by resistivity and magnetization measurements. Two additional
finely pulverized single crystal samples ($x = 0.075, 0.1$), grown by the floating zone method
utilizing image furnace, were also used to increase coverage of the
phase space under the study. Neutron powder diffraction measurements
were carried out at the GEM diffractometer at the ISIS facility at
the Rutherford Appleton Laboratory in the UK and at the NPDF
diffractometer at Los Alamos Neutron Scattering Center at Los
Alamos National Laboratory. The samples, of approximately 6 grams
each, were loaded into extruded vanadium containers and sealed
under He atmosphere. The data were collected for all the samples
at a consistent set of 7 temperatures between 10~K and 550~K using
a closed cycle He refrigerator. The data were processed to
obtain PDFs~\cite{peter;jac03} using the program PDFgetN~\cite{peter;jac00}
by a sine Fourier transform of the total scattering structure
function $F(Q)$ up to a value of $Q_{max}$ of
35~\AA$^{-1}$. This high $Q_{max}$, coupled with the good statistics
from GEM and NPDF, result in high-quality PDFs with minimal spurious
low-$r$ ripples and negligible termination ripples, as evident in
Fig.~\ref{fig:grdata}.
%%%%%%%%%%%%%%%%%%%%%%%%%%%%%%%%%%%%
%        FIGURE #1
%%%%%%%%%%%%%%%%%%%%%%%%%%%%%%%%%%%%
%%%%%%%%%%%%%%%%%%%%%%%%%%%%%%%%%%%
%
%%%%%%%%%%%%%%%%%%%%%%%%%%%%%%%%%%%
The PDF analysis reported here involves both direct data evaluation
and structural modeling using the program PDFFIT~\cite{proff;jac99}.
Results of complementary average crystal structure modeling, carried
out using the program GSAS~\cite{larso;unpub87}, are also presented.
All refinements were carried out using the O or O$^\prime$ structural
models in the Pbnm space-group with isotropic displacement
parameters~\cite{proff;prb99}. Our analysis does not address the
ordering of localized charges such as observed in the charge ordered 
state at $x=0.5$~\cite{radae;prb97ii}.

\begin{figure}
\includegraphics[angle=0,width=0.45\textwidth]{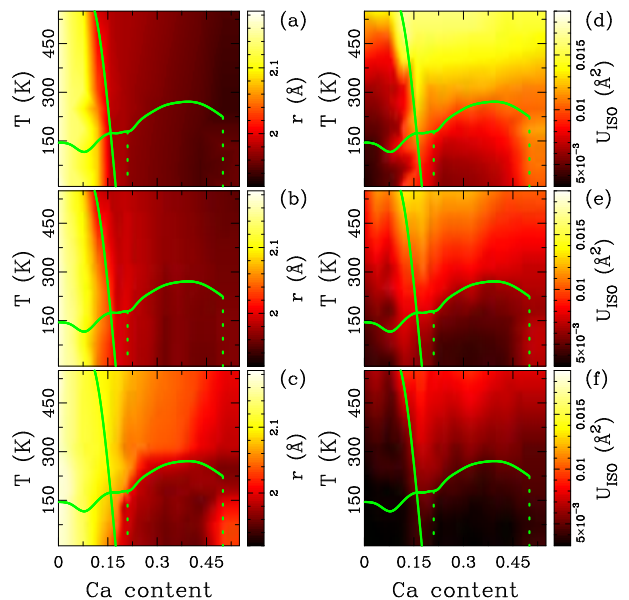}
\caption{\label{fig:jtdist} Contour plot of the JT distortion (long
Mn-O distance) in (x,T) parameter space as obtained from (a) Rietveld
analysis, and PDF analyses over (b) 20~\AA\ and (c) 6~\AA\ ranges,
respectively. Contour plot of the isotropic displacement parameter
of oxygen, U$_{iso}$(O2) in $Pbnm$ setting, as obtained from (d)
Rietveld analysis, and PDF analyses over (e) 20~\AA\ and (f) 6~\AA\
ranges. In all the panels the solid curves indicate T$_{JT}$ and
T$_{c}$ phase lines, while dotted vertical lines indicate IM phase
boundaries, as determined from the sample characterization measurements.}
\end{figure}

\begin{figure}
\includegraphics[angle=0,width=0.38\textwidth]{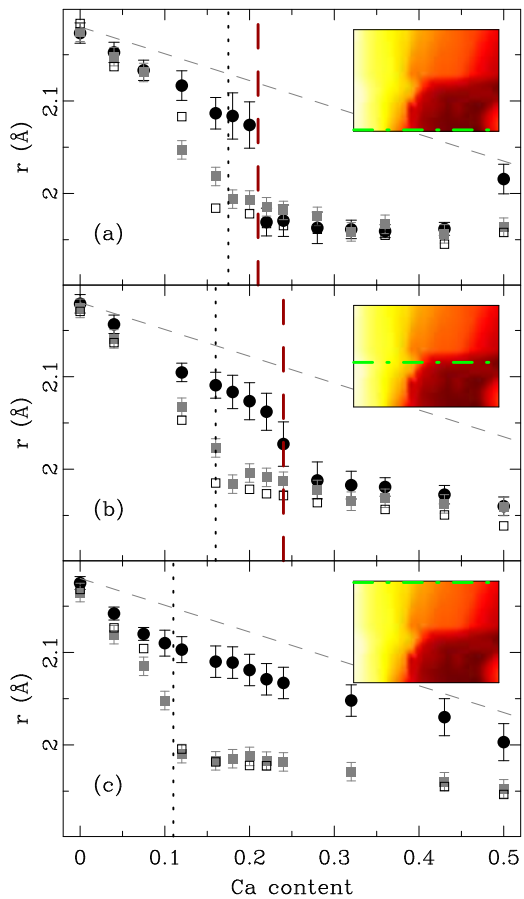}
\caption{\label{fig:jtdistcuts} Length of the longest Mn-O distance
in the MnO$_{6}$ octahedron vs doping at (a) 10~K, (b) 250~K, and
(c) 550~K. Open squares denote Rietveld result, solid squares
and solid circles show results of PDF refinements over 20~\AA~
and 6~\AA~ ranges, respectively. Sloping dashed lines denote
Vegard's law~\cite{vegar;zp21} behavior for Mn-O long bond
interpolating between the values for \lmo~ and \cmo . Dashed
vertical lines mark the IM transition, while the dotted vertical
lines denote the orthorhombic to pseudocubic phase transition. The
insets reproduce Fig.~\ref{fig:jtdist}(c) showing the temperature
of the data shown in the panel. Symbols indicate measured compositions.}
\end{figure}

The results of the average crystal structure evaluation is 
summarized in Figs.~\ref{fig:jtdist}(a) and~\ref{fig:jtdistcuts}.
%%%%%%%%%%%%%%%%%%%%%%%%%%%%%%%%%%%%
%        FIGURE #2
%%%%%%%%%%%%%%%%%%%%%%%%%%%%%%%%%%%%
%%%%%%%%%%%%%%%%%%%%%%%%%%%%%%%%%%%
%
%%%%%%%%%%%%%%%%%%%%%%%%%%%%%%%%%%%
In the orthorhombic O-phase~\cite{wolla;pr55}, the \emph{local} JT
distortion amplitude (the length of the JT long-bond) is the same as
the average long range ordered value. The distortion is constant
with temperature for fixed Ca content, but decreases linearly with
increasing Ca content. Upon crossing into the pseudo-cubic
O$^\prime$ phase, the average JT distortion disappears abruptly
(Fig~\ref{fig:jtdist}(a) and~\ref{fig:jtdistcuts}). However, this
effect is accompanied by the anomalous increase of the isotropic
thermal parameters on oxygen sites, Fig~\ref{fig:jtdist}(d).  This
is consistent with the understanding that the pseudo-cubic
O$^\prime$ phase in the insulating state consists of orbitally
disordered, JT distorted, octahedra~\cite{qiu;prl05,rodri;prb98},
and that this picture can be extended to finite doping.  In the
ferromagnetic metallic (FM) phase, the Rietveld refined JT long-bond
disappears, but there is no significant enlargement of the refined
oxygen thermal parameter showing that the local JT long-bond is also
absent~\cite{billi;prl96}. Structural refinements to the PDF data
over a wide range of $r$ ($r_{max}=20$~\AA ) mimic the Rietveld
results rather closely (Fig.~\ref{fig:jtdist}(b) and (e)).  This
shows that the average structure result is already obtained for a
PDF refinement over a 20~\AA\ range, suggesting that the size of
local orbital ordering correlations is limited to this range.

The size and shape of the local MnO$_6$ octahedra can be obtained by
fitting the PDF over a narrow $r$-range of 6~\AA . The length of the
local long Mn-O bond has been obtained for all compositions and
temperatures studied, and is shown as a contour plot in
Fig.~\ref{fig:jtdist}(c). The color-scale has been set such that a
fully shortened long-bond of 1.96~\AA\ shows up as black. The
presence of color therefore indicates a finite local JT distortion.
There is a striking resemblance between the contour plot of the
local JT long-bond in Fig.~\ref{fig:jtdist}(c) and the electronic
phase diagram of this manganite that is superimposed. The position
of the phase transition lines were verified from magnetization and
resistivity measurements of the samples used in this study.

Firstly we note that the \emph{local} JT distortion is present in
the entire insulating part of the phase diagram, but that it is
effectively removed for the metallic compositions at lowest
temperatures. Secondly, it is seen that the \emph{magnitude} of the
local JT distortion has a relatively strong doping dependence at
lower Ca-concentrations, with the bond length vs. concentration
curve flattening at higher Ca doping levels.

Selected constant temperature cuts are shown in
Fig.~\ref{fig:jtdistcuts}(a)-(c), for 550~K, 310~K, and 10~K
respectively.
%%%%%%%%%%%%%%%%%%%%%%%%%%%%%%%%%%%%
%        FIGURE #3
%%%%%%%%%%%%%%%%%%%%%%%%%%%%%%%%%%%%
%%%%%%%%%%%%%%%%%%%%%%%%%%%%%%%%%%%%
%
%%%%%%%%%%%%%%%%%%%%%%%%%%%%%%%%%%%%
The square symbols show the behavior of the average structure.  The
JT long-bond decreases with doping in the orbitally ordered O-phase,
but then abruptly shortens at the structural phase transition,
indicated by the dotted line in the figure.  In contrast, the local
JT bond is insensitive to the structural transition, but disappears
abruptly when the sample goes through the insulator-metal (IM)
transition.  This is consistent with the widely held current
view~\cite{billi;prl96, booth;prl98}.  What is  less expected is the
observation that the {\it length} of the local JT-long bond shortens
with increasing doping in the insulating state. In a single-site
small polaron picture the insulating state consists of distinct
Mn$^{3+}$ and Mn$^{4+}$ sites.  The 4+ sites are presumed to have
regular, undistorted, MnO$_6$ octahedra and the 3+ sites to be JT
distorted with 2.16~\AA\ long bonds, as in the undoped LaMnO$_3$
endmember, and this was supported by experimental
evidence~\cite{louca;prb97,booth;prl98,billi;prb00}. We investigate
these refinement results in greater detail below.

\begin{figure}
\includegraphics[angle=-90,width=0.38\textwidth]{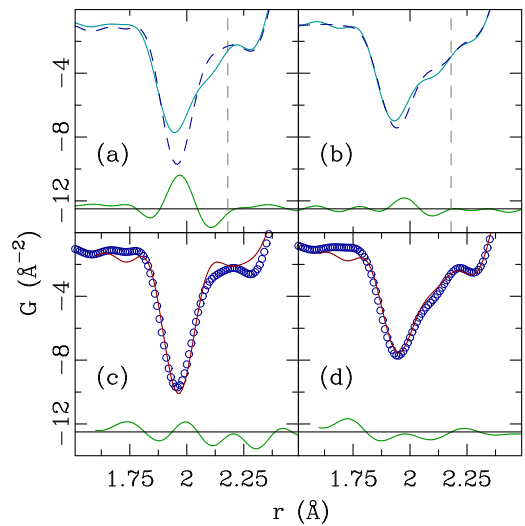}
\caption{\label{fig:jtdistfit} Experimental PDFs, G(r), of x=0.18
(solid line) and x=0.22 (dashed line) samples at (a) 10~K and (b)
190~K, with difference curves underneath. Both datasets are in the
insulating phase in (b) but straddle the IM boundary in (a). A JT
long bond appears in the insulating phase but is shorter than that
in the undoped endmember, indicated by the vertical dashed lines.
(c) and (d) show representative r$_{max}$=6~\AA~ PDF fits (solid lines)
to the 10~K data (open circles) shown in (a). Difference curves are
offset below.}
\end{figure}

Fig.~\ref{fig:jtdistfit} shows some representative PDF data
bracketing the IM transition.
%%%%%%%%%%%%%%%%%%%%%%%%%%%%%%%%%%%%
%        FIGURE #4
%%%%%%%%%%%%%%%%%%%%%%%%%%%%%%%%%%%%
%%%%%%%%%%%%%%%%%%%%%%%%%%%%%%%%%%%%
%
%%%%%%%%%%%%%%%%%%%%%%%%%%%%%%%%%%%%
We show the nearest neighbor doublet of the PDF from \lcmo\ with
$x=0.18$ and $x=0.22$.  The former sample remains in the insulating
state at all temperatures; the latter is a ferromagnetic metal 
below $\sim 180~K$, and paramagnetic insulator at higher temperature.
Thus, a comparison of the curves in Fig.~\ref{fig:jtdistfit}(b)
shows the effect of changing composition without crossing the IM
transition.  The changes are very small.  In
Fig.~\ref{fig:jtdistfit}(a) the MI is crossed at constant
temperature.  The changes to the nearest-neighbor doublet on
crossing the MI transition are evident with a sharp, single-valued
peak giving way to a peak with a well-defined shoulder on the
high-$r$ side.  This is the appearance of the JT long-bond in the
insulating phase.  Notable, however, is the absence of intensity at
the position of the undoped JT-long bond at $r=2.16$~\AA , indicated
by a vertical dashed line in the figure.  This clearly shows the
absence of any fully distorted JT octahedra in the structure. The
difference curve in Fig.~\ref{fig:jtdistfit}(a) also shows the
appearance of intensity on the low-$r$ side of the shortest bond in
the insulating phase.  Fig.~\ref{fig:jtdistfit}(c) and (d) shows the
model fits to the metallic and insulating PDFs, giving a sense of
the quality of fits to the local PDF that were obtained in our
determination of the phase diagram in Figs.~\ref{fig:jtdist} and
\ref{fig:jtdistcuts}.

 The main result of the current study is the clear
demonstration of an absence of any Mn-O long-bonds of the fully JT
distorted 2.16~\AA\ length in the doped samples, as would be
expected in a simple-minded small-polaron picture. Our result
disagrees with the canonical understanding from earlier experimental
studies~\cite{louca;prb97,booth;prl98,billi;prb00}, though is
consistent with the observations of a number of less well known
studies~\cite{hibbl;jpcm99,bindu;jpcm05}. The observation could be
explained in a homogeneous picture where the electron density is
uniformly distributed over all Mn sites in the insulating state. In
this case, the electron density in the orbitals, and therefore the
driving force for the JT distortion, decreases continuously with
increasing hole doping.   It is harder to explain in a small-polaron
picture where localized Mn$^{4+}$ sites exist in a fully-distorted
Mn$^{3+}$ background.  In this scenario, one source of long-bond
shortening could be elastic strain in the lattice because of the
disordered polarons.  This is unlikely to explain our observation.
The energy scale for octahedral rotations is much smaller than for
stretching the covalent bonds~\cite{iliev;prb98} and most elastic
strain accommodation is expected to occur in the octahedral
rotations.  This is indeed the case when orbital disorder sets in at
constant doping as a function of temperature~\cite{qiu;prl05}. Also,
we note that there is a negative deviation from Vegard's
law~\cite{vegar;zp21}, indicated by the dashed line in
Fig.~\ref{fig:jtdistcuts}, in the average length of the long-bond.
Elastic relaxation in covalently bonded alloys analogous to the
manganites results in a characteristic
Z-plot~\cite{mikke;prl82,jeong;prb01} of bond-length vs. composition
that is clearly not seen here. A principle strain relaxation mode
through octahedral rotations is also indicated by enlarged $U_{iso}$
values in the insulating region of the phase diagram, as evident in
Fig.~\ref{fig:jtdist}(d). We also note that oxygen K-edge
spectroscopy indicates an increasing importance of oxygen ligand
hole states with doping~\cite{elp;prb99,zampi;prb98}, indicating a
clear change in the local electronic structure induced by doping.

It is not possible from our current data to tell unambiguously
whether or not the sample is phase separated into metallic and
insulating regions.  Phase separation would imply a transfer of
intensity from the position of the long-bond to that of the
short-bond.  What we show here is that the length of the long-bond
decreases with doping. The long-bond starts to overlap with the
short-bond at higher doping making an extraction of the integrated
area of the peak uncertain. However, we note that modeling the local
structure with a homogeneous model in Pbnm setting as we have done,
with 4-short and 2-long bonds at all doping levels in the insulating
region, results in rather good fits as evident in
Figs.~\ref{fig:grdata} and ~\ref{fig:jtdistfit}(d). Enlarged oxygen
$U_{iso}$ values around $x=0.18$ are evident in
Fig.~\ref{fig:jtdist}(d)-(f) where phase separation has been
observed~\cite{pissa;prb05}.

In summary, we have presented the most complete and highest quality
neutron PDF measurements of the cubic \lcmo\ phase over a wide range
of temperature and doping up to 50\%.  The data clearly indicate
that the {\it local} JT distortion decreases in magnitude smoothly
and continuously with doping in the insulating phase.  This cannot
be easily explained in the canonical small-polaron picture and
calls for a theoretical explanation.

%%%%%%%%%%%%%%%%
E.B. acknowledges discussions with S.D. Mahanti, P.M. Duxbury, and
T.A. Kaplan, and help from P. Juh\'{a}s. This work was supported by
the NSF under grant DMR-0304391, and benefited from usage of NPDF 
at LANSCE at LANL. Argonne and Los Alamos National Laboratories are 
operated under DOE Contracts No. DE-AC02-06CH11375 and DE-AC52-06NA25396 
respectively.
%%%%%%%%%%%%%%%%


\begin{thebibliography}{10}

\bibitem{milli;n98}
A.~J. Millis,
\newblock Nature {\bf 392}, 147 (1998).

\bibitem{komiy;prl05}
S.~Komiya, H.~D.~Chen, S.~C. Zhang, and Y.~Ando,
\newblock Phys. Rev. Lett. {\bf 94}, 207004 (2005).

\bibitem{hanag;n04}
T.~Hanaguri, C.~Lupien, Y.~Kohsaka, D.~H. Lee, M.~Azuma, M.~Takano, H.~Takagi,
  and J.~C. Davis,
\newblock Nature {\bf 430}, 1001 (2004).

\bibitem{tranq;n95}
J.~M. Tranquada, B.~J. Sternlieb, J.~D. Axe, Y.~Nakamura, and S.~Uchida,
\newblock Nature {\bf 375}, 561 (1995).

\bibitem{fath;s99}
M.~Fath, S.~Freisem, A.~A. Menovsky, Y.~T. Y, J.~Aarts, and J.~A. Mydosh,
\newblock s , 1540 (1999).

\bibitem{uehar;n99}
M.~Uehara, S.~Mori, C.~H. Chen, and {S.-W. Cheong},
\newblock Nature {\bf 399}, 560 (1999).

\bibitem{becke;prl02}
T.~Becker, C.~Streng, Y.~Luo, V.~Moshnyaga, B.~Damaschke, N.~Shannon, and
  K.~Samwer,
\newblock Phys. Rev. Lett. {\bf 89}, 237203 (2002).

\bibitem{loudo;prl06}
J.~C. Loudon and P.~A. Midgley,
\newblock Phys. Rev. Lett. {\bf 96}, 027214 (2006).

\bibitem{dagot;s05}
E.~Dagotto,
\newblock Science {\bf 309}, 257 (2005).

\bibitem{milwa;n05}
G.~C. Milward, M.~J. {Calderon}, and P.~B. Littlewood,
\newblock Nature {\bf 433}, 607 (2005).

\bibitem{akiya;apl01}
R.~Akiyama, H.~Tanaka, T.~Matsumoto, and T.~Kawai,
\newblock Appl. Phys. Lett. {\bf 79}, 4378 (2001).

\bibitem{matsu;pc03}
K.~Matsuba, H.~Sakata, T.~Mochiku, K.~Hirata, and N.~Nishida,
\newblock Physica C {\bf 388}, 281 (2003).

\bibitem{mcelr;s05}
K.~McElroy, J.~Lee, J.~A. Slezak, D.~H. Lee, H.~Eisaki, S.~Uchida, and J.~C.
  Davis,
\newblock Science {\bf 309}, 1048 (2005).

\bibitem{billi;prl96}
S.~J.~L. Billinge, R.~G. DiFrancesco, G.~H. Kwei, J.~J. Neumeier, and J.~D.
  Thompson,
\newblock Phys. Rev. Lett. {\bf 77}, 715 (1996).

\bibitem{booth;prl98}
C.~H. Booth, F.~Bridges, G.~H. Kwei, J.~M. Lawrence, A.~L. Cornelius, and J.~J.
  Neumeier,
\newblock Phys. Rev. Lett. {\bf 80}, 853 (1998).

\bibitem{radae;prb96i}
P.~G. Radaelli, M.~Marezio, H.~Y. Hwang, S.~W. Cheong, and B.~Batlogg,
\newblock Phys. Rev. B {\bf 54}, 8992 (1996).

\bibitem{louca;prb97}
D.~Louca, T.~Egami, E.~L. Brosha, H.~{R\"{o}der}, and A.~R. Bishop,
\newblock Phys. Rev. B {\bf 56}, R8475 (1997).

\bibitem{billi;prb00}
S.~J.~L. Billinge, T.~Proffen, V.~Petkov, J.~L. Sarrao, and S.~Kycia,
\newblock Phys. Rev. B {\bf 62}, 1203 (2000).

\bibitem{proff;apa01ii}
T.~Proffen and S.~J.~L. Billinge,
\newblock Appl. Phys. A {\bf 74}, 1770 (2002).

\bibitem{dagot;prpl01}
E.~Dagotto, T.~Hotta, and A.~Moreo,
\newblock Phys. Rep. {\bf 344}, 1 (2001).

\bibitem{massa;jmmm01}
N.~E. Massa, H.~C.~N. Tolentino, H.~Salva, J.~A. Alonso, M.~J. Martinez-Lope,
  and M.~T. Casais,
\newblock J. Magn. Magn. Mater. {\bf 233}, 91 (2001).

\bibitem{dabro;jssc89}
B.~Dabrowski, R.~Dybzinski, Z.~Bukowski, O.~Chmaissem, and J.~D. Jorgensen,
\newblock J. Solid State Chem. {\bf 146}, 448 (1989).

\bibitem{peter;jac03}
P.~F. Peterson, {E. S. Bo\v zin}, T.~Proffen, and S.~J.~L. Billinge,
\newblock J. Appl. Crystallogr. {\bf 36}, 53 (2003).

\bibitem{peter;jac00}
P.~F. Peterson, M.~Gutmann, T.~Proffen, and S.~J.~L. Billinge,
\newblock J. Appl. Crystallogr. {\bf 33}, 1192 (2000).

\bibitem{proff;jac99}
T.~Proffen and S.~J.~L. Billinge,
\newblock J. Appl. Crystallogr. {\bf 32}, 572 (1999).

\bibitem{larso;unpub87}
A.~C. Larson and R.~B. {Von Dreele},
\newblock General structure analysis system,
\newblock Report No. LAUR-86-748, Los Alamos National Laboratory, Los Alamos,
  NM 87545, 1987.

\bibitem{proff;prb99}
T.~Proffen, R.~G. DiFrancesco, S.~J.~L. Billinge, E.~L. Brosha, and G.~H. Kwei,
\newblock Phys. Rev. B {\bf 60}, 9973 (1999).

\bibitem{radae;prb97ii}
P.~G. Radaelli, D.~E. Cox, M.~Marezio, and S.-W. Cheong,
\newblock Phys. Rev. B {\bf 55}, 3015 (1997).

\bibitem{wolla;pr55}
E.~O. Wollan and W.~C. Koehler,
\newblock Phys. Rev. {\bf 100}, 545 (1955).

\bibitem{qiu;prl05}
X.~Qiu, {Th.~Proffen}, J.~F. Mitchell, and S.~J.~L. Billinge,
\newblock Phys. Rev. Lett. {\bf 94}, 177203 (2005).

\bibitem{rodri;prb98}
J.~Rodriguez-Carvajal, M.~Hennion, F.~Moussa, A.~H. Moudden, L.~Pinsard, and
  A.~Revcolevschi,
\newblock Phys. Rev. B {\bf 57}, R3189 (1998).

\bibitem{vegar;zp21}
L.~Vegard,
\newblock Z. Phys. {\bf 5}, 17 (1921).

\bibitem{hibbl;jpcm99}
S.~J. Hibble, S.~P. Cooper, A.~C. Hannon, I.~D. Fawcett, and M.~Greenblatt,
\newblock J. Phys: Condens. Matter {\bf 11}, 9221 (1999).

\bibitem{bindu;jpcm05}
R.~Bindu, S.~K. Pandey, A.~Kumar, S.~Khalid, and A.~V. Pimpale,
\newblock J. Phys: Condens. Matter {\bf 17}, 6393 (2005).

\bibitem{iliev;prb98}
M.~N. Iliev, M.~V. Abrashev, H.-G. Lee, V.~N. Popov, Y.~Y. Sun, C.~Thomsen,
  R.~L. Meng, and C.~W. Chu,
\newblock Phys. Rev. B {\bf 57}, 2872 (1998).

\bibitem{mikke;prl82}
J.~C. Mikkelsen and J.~B. Boyce,
\newblock Phys. Rev. Lett. {\bf 49}, 1412 (1982).

\bibitem{jeong;prb01}
{I.-K.~Jeong}, F.~Mohiuddin-Jacobs, V.~Petkov, S.~J.~L. Billinge, and S.~Kycia,
\newblock Phys. Rev. B {\bf 63}, 205202 (2001).

\bibitem{elp;prb99}
J.~van Elp and A.~Tanaka,
\newblock Phys. Rev. B {\bf 60}, 5331 (1999).

\bibitem{zampi;prb98}
G.~Zampieri, F.~Prado, A.~Caneiro, J.~Bri\'{a}tico, M.~T. Causa, M.~Tovar,
  B.~Alascio, M.~Abbate, and E.~Morikawa,
\newblock Phys. Rev. B {\bf 58}, 3755 (1998).

\bibitem{pissa;prb05}
M.~Pissas, I.~Margiolaki, G.~Papavassiliou, D.~Stamopoulos, and D.~Argyriou,
\newblock Phys. Rev. B {\bf 72}, 064425 (2005).

\end{thebibliography}
\end{document}